\documentclass[twocolumn]{aastex631}

\usepackage{amsmath}

\definecolor{fuchsia}{rgb}{0.54, 0.17, 0.89}
\definecolor{azure}{rgb}{0.0, 0.5, 1.0}
\definecolor{pgreen}{rgb}{0.12, 0.3, 0.17}
\definecolor{alizarin}{rgb}{0.82, 0.1, 0.26}

\newcommand{\kms}{{\rm km~s^{-1}}}
\newcommand{\oi}{$\textrm{O}~\textsc{i}_{\lambda1304}$}
\newcommand{\oii}{$[\textrm{O}~\textsc{ii}]_{\lambda 3727}$}
\newcommand{\oiii}{[\textrm{O}~\textsc{iii}]}
\newcommand{\oiiir}{$[\textrm{O}~\textsc{iii}]_{\lambda5007}$}

\newcommand{\oiiiuv}{$\textrm{O}~\textsc{iii}]_{\lambda1661}$}
\newcommand{\oiiif}{$[\textrm{O}~\textsc{iii}]_{\lambda4363}$}

\newcommand{\niii}{$\textrm{N}~\textsc{iii}]_{\lambda1747}$}
\newcommand{\niv}{$\textrm{N}~\textsc{iv}]_{\lambda1488}$}
\newcommand{\nv}{$\textrm{N}~\textsc{v}_{\lambda1240}$}
\newcommand{\neiii}{[\textrm{Ne}~\textsc{iii}]}
\newcommand{\neiv}{[$\textrm{Ne}~\textsc{iv}]_{\lambda2423}$}
\newcommand{\nev}{[$\textrm{Ne}~\textsc{v}]_{\lambda3346}$}

\newcommand{\cii}{[\textrm{C}~\textsc{ii}]}
\newcommand{\ciii}{$\textrm{C}~\textsc{iii}]_{\lambda1908}$}
\newcommand{\civ}{$\textrm{C}~\textsc{iv}_{\lambda1548}$}

\newcommand{\heii}{\textrm{He}~\textsc{ii}}
\newcommand{\heiiuv}{$\textrm{He}~\textsc{ii}_{\lambda1640}$}

\newcommand{\ly}{${\rm Ly\alpha}$}

\newcommand{\hb}{${\rm H\beta}$}
\newcommand{\hg}{${\rm H\gamma}$}
\newcommand{\hd}{${\rm H\delta}$}

\newcommand{\simgt}{\,\rlap{\lower 3.5 pt \hbox{$\mathchar \sim$}} \raise
1pt \hbox {$>$}\,}
\newcommand{\simlt}{\,\rlap{\lower 3.5 pt \hbox{$\mathchar \sim$}} \raise
1pt \hbox {$<$}\,}

\newcommand{\Z}{{\rm [Z/H]}}

\newcommand{\id}{CANUCS-LRD-z8.6}
\newcommand{\rb}{1.5}
\newcommand{\rberr}{$\rb_{-0.2}^{+0.3}$}
\newcommand{\esc}{11}
\newcommand{\escerr}{$\esc \pm 3$}
\newcommand{\fwint}{2200}
\newcommand{\fwinterr}{$\fwint\pm280$}
\newcommand{\fw}{1540}
\newcommand{\fwerr}{$\fw\pm260$}

\newcommand{\dod}{$1.8_{-0.6}^{+3.0}$} % 
\newcommand{\nphot}{$8$} % 
\newcommand{\nexp}{$2.8_{-1.7}^{+1.5}$}

\submitjournal{ApJ}

\shorttitle{Peculiar \ly-emitting AGN at $z=8.63$}
\shortauthors{Morishita et al.}

\graphicspath{{./}{figures/}}
\begin{document}

\title{A Nitrogen-rich AGN Powering a Large Ionizing Bubble at z=8.63}

\correspondingauthor{Takahiro Morishita}
\email{takahiro@ipac.caltech.edu}

\author[0000-0002-8512-1404]{Takahiro Morishita}
\affiliation{IPAC, California Institute of Technology, MC 314-6, 1200 E. California Boulevard, Pasadena, CA 91125, USA}
\affiliation{Astronomical Institute, Tohoku University, 6-3, Aramaki, Aoba, Sendai, Miyagi 980-8578, Japan}

\author[0000-0001-9935-6047]{Massimo Stiavelli}
\affiliation{Space Telescope Science Institute, 3700 San Martin Drive, Baltimore, MD 21218, USA}
% \affiliation{The William H. Miller III, Dept. of Physics \& Astronomy, Johns Hopkins University, Baltimore, MD 21218, USA}
% \affiliation{Dept. of Astronomy, University of Maryland, College Park, MD 20742, USA}

% \author{et al.}

\author[0000-0002-3407-1785]{Charlotte A. Mason}
\affiliation{Cosmic Dawn Center (DAWN)}
\affiliation{Niels Bohr Institute, University of Copenhagen, Jagtvej 128, DK-2200 Copenhagen N, Denmark}

\author[0000-0002-9909-3491]{Roberta Tripodi}
\affiliation{Faculty of Mathematics and Physics, University of Ljubljana, 19 Jadranska ulica, Ljubljana, 1000, Slovenia}
\affiliation{IFPU, Institute for Fundamental Physics of the Universe, Via Beirut 2, Trieste, 34151, Italy}

\author[0000-0003-1564-3802]{Marco Chiaberge}
\affiliation{Space Telescope Science Institute for the European Space Agency (ESA), ESA Office, 3700 San Martin Drive, Baltimore, MD 21218, USA}
\affiliation{The William H. Miller III Department of Physics and Astronomy, Johns Hopkins University, Baltimore, MD 21218, USA}

\author[0000-0003-2497-6334]{Stefan Schuldt}
\affiliation{Dipartimento di Fisica, Università degli Studi di Milano, Via Celoria 16, I-20133 Milano, Italy}
\affiliation{INAF - IASF Milano, via A. Corti 12, I-20133 Milano, Italy}

\author[0000-0002-4201-7367]{Chris J. Willott}
\affiliation{NRC Herzberg, 5071 West Saanich Rd, Victoria, BC V9E 2E7, Canada}

\author[0000-0003-3817-8739]{Yechi Zhang}
\affiliation{IPAC, California Institute of Technology, MC 314-6, 1200 E. California Boulevard, Pasadena, CA 91125, USA}

%% Note that the \and command from previous versions of AASTeX is now
%% depreciated in this version as it is no longer necessary. AASTeX 
%% automatically takes care of all commas and "and"s between authors names.

%% AASTeX 6.31 has the new \collaboration and \nocollaboration commands to
%% provide the collaboration status of a group of authors. These commands 
%% can be used either before or after the list of corresponding authors. The
%% argument for \collaboration is the collaboration identifier. Authors are
%% encouraged to surround collaboration identifiers with ()s. The 
%% \nocollaboration command takes no argument and exists to indicate that
%% the nearby authors are not part of surrounding collaborations.

%% Mark off the abstract in the ``abstract'' environment. 
\begin{abstract}
We report the detection of \ly\ in \id, a recently discovered active galactic nucleus (AGN) at $z=8.63$ by \citet{tripodi24}, in new NIRSpec/MSA G140H/F070LP observations. We detect broad \ly\ emission (FWHM\,=\fwerr\,km/s) near the systemic velocity, which suggests a large ionizing bubble considering that the universe is almost fully neutral at the redshift. Through \ly\ line-shape modeling assuming a Str{\"o}mgren sphere, we find a large bubble radius, $R_b=$\,\rberr\,pMpc, and a moderately high \ly\ escape fraction, $f_{\rm esc}=$\,\escerr\,\%. The intrinsic line width is inferred to be broad (\fwinterr\,km/s), likely originating in the broad-line region. Existing data indicate that \id\ is within a mild overdensity, $\delta = $\,{\dod}, suggesting that other galaxies in its proximity might have contributed to the formation of the bubble. The high \niv/\civ\ and \niv/\oiiiuv\ line ratios measured in existing NIRSpec/PRISM data indicate nitrogen enrichment in this metal-poor, low-luminosity AGN. The spectroscopic features are overall similar to other nitrogen-rich galaxies discovered in the literature, such as GN-z11 and GHZ2/GLASSz12. This suggests that \id\ may represent one of the evolutionary phases of those nitrogen-rich galaxies. 
\end{abstract}

%% Keywords should appear after the \end{abstract} command. 
%% The AAS Journals now uses Unified Astronomy Thesaurus concepts:
%% https://astrothesaurus.org
%% You will be asked to selected these concepts during the submission process
%% but this old "keyword" functionality is maintained in case authors want
%% to include these concepts in their preprints.
\keywords{}
%Classical Novae (251) --- Ultraviolet astronomy(1736) --- History of astronomy(1868) --- Interdisciplinary astronomy(804)}

%% From the front matter, we move on to the body of the paper.
%% Sections are demarcated by \section and \subsection, respectively.
%% Observe the use of the LaTeX \label
%% command after the \subsection to give a symbolic KEY to the
%% subsection for cross-referencing in a \ref command.
%% You can use LaTeX's \ref and \label commands to keep track of
%% cross-references to sections, equations, tables, and figures.
%% That way, if you change the order of any elements, LaTeX will
%% automatically renumber them.
%%
%% We recommend that authors also use the natbib \citep
%% and \citet commands to identify citations.  The citations are
%% tied to the reference list via symbolic KEYs. The KEY corresponds
%% to the KEY in the \bibitem in the reference list below. 

\section{Introduction} \label{sec:intro}
The James Webb Space Telescope (JWST; \citealt{gardner23}) has started revolutionizing our understanding of galaxies and stellar populations in the first billion years. With its exquisite sensitivity, deep surveys have identified new sources well beyond the previous limit \citep[e.g.,][]{curtis-lake23,bunker23,castellano24,carniani24,naidu25}, now up to $z\sim14$. The identification of sources at high redshifts and their almost immediate spectroscopic characterization have given us great insights into the formation of the first galaxies, black holes, and stars \citep[e.g.,][]{robertson23,dekel23,ziparo23,inayoshi24}.

Of particular interest are unusual nucleosynthesis patterns (to our local standard) discovered in intensely star-forming objects at $z>6$. For example, sensitive spectroscopy with JWST/NIRSpec revealed over-abundance of nitrogen in GNz11 \citep{bunker23,cameron23,senchyna23,charbonnel23} and in other galaxies \citep{marques-chaves24,schaerer24,topping24,ji24,isobe23b}, over- and under-abundance of Carbon \citep{stiavelli25,jones23,hsiao23,deugenio24}. In particular, those with nitrogen-rich objects are characterized with a high specific star formation rate, extremely compact morphology, and in a few cases broad emission lines. All of these characteristics point to intense star formation within the scale of $\simlt100$\,pc, likely driven by massive stars of low-metallicity \citep{schaerer24,topping24}. Probing different elemental abundances can reveal information on the detailed galaxy formation process, as stars of different masses, and hence different lifetimes, are the main producers of different elements \citep[e.g.,][]{stiavelli25}.

However, the prevalence of active galactic nuclei (AGN) at high redshift \citep[e.g.,][]{harikane23c} complicates the story, making it less straightforward to interpret the observed properties. An example is seen in the aforementioned GN-z11, which revealed several high-ionization rest-frame UV lines, including \niv, \civ, and \ciii\ \citep[][]{bunker23}. \citet{maiolino24} reported high gas densities ($\simgt10^9$\,cm$^{-3}$) and the AGN characteristic \neiv\ and \cii$_{\lambda 1335}$ lines in the same source, and favored the presence of an AGN over other non-AGN origins (e.g., Wolf-Rayet stars). Similarly, \citet{castellano24} reported the detection of \niv, \civ, \heiiuv, \oiiiuv, and \ciii\ lines in GHZ2/GLASS-z12. While the high ionization \neiv\ line was not detected in the spectrum of GHZ2/GLASS-z12, the authors showed that the source is still compatible with AGN from various emission line diagnostics. 

Emission line diagnostics used in those studies are often established with local samples, and thus may not serve as useful as for local studies --- due to the presence of hard ionizing sources, e.g., massive stars in extremely metal-poor, high gas density environments \citep[e.g.,][]{mazzolari24,cleri25}. The lack of wavelength coverage for \hb\ at $z>9.3$ by NIRSpec makes the solid confirmation of broad-line AGN challenging in both examples above. In this regard, there exists a need for deep spectroscopic coverage at $>5\,\mu$m by MIRI \citep[e.g.,][]{zavala25} or detailed studies of analogs at slightly lower redshifts \citep[e.g.,][]{rojaz-ruiz25}.

\begin{figure*}
\centering
	\includegraphics[width=0.9\textwidth]{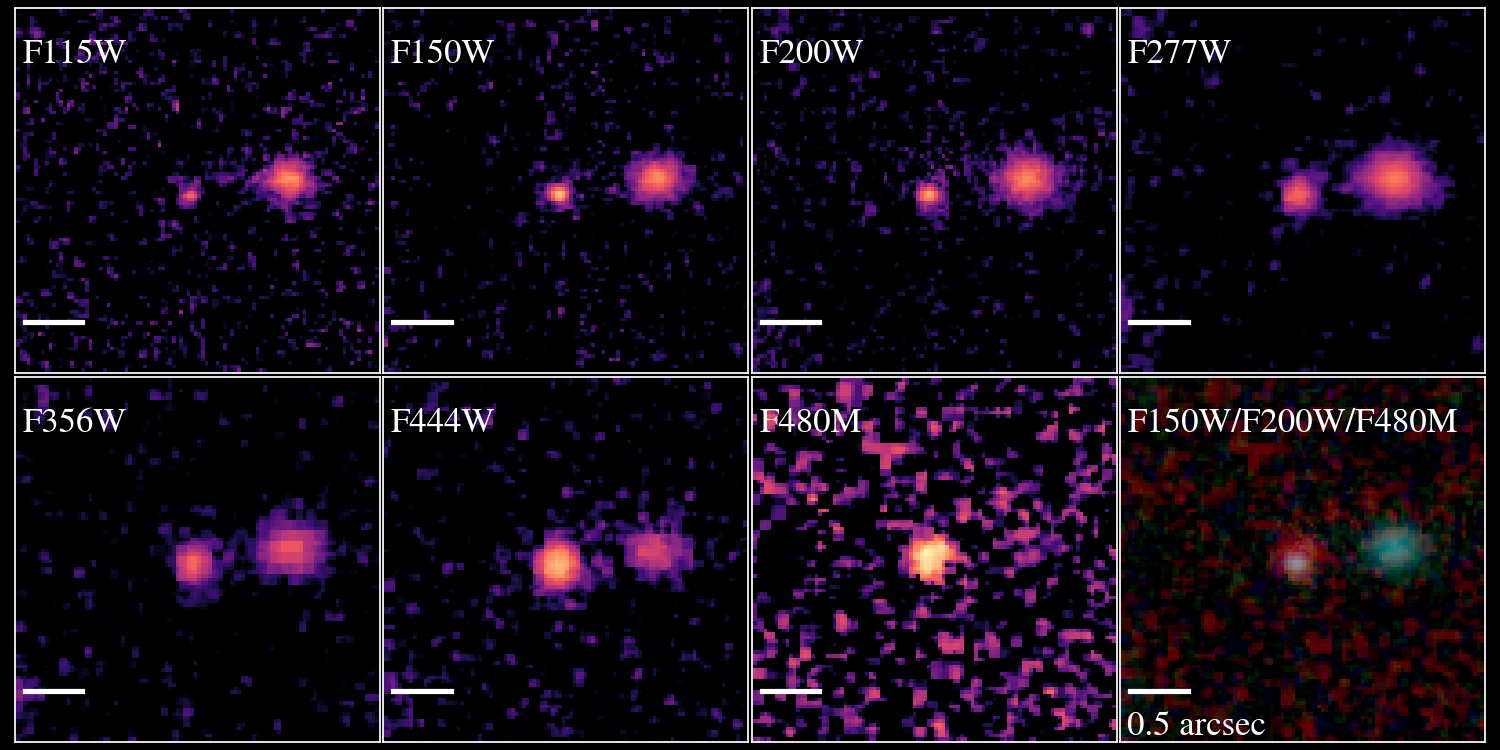}
	\caption{
    Postage stamps of \id\ in JWST/NIRCam filters in the cutout size of $3.\!''2$. A pseudo color image (F150W/F200W/F480M for blue, green, and red) is also shown.
    The extended object to the west is a foreground source at $z_{\rm phot}\sim0.4$.
    }
\label{fig:stamp}
\end{figure*}

In this paper, we present a new observation of a previously reported AGN at $z=8.63$, \id, in the sightline of {MACS~J1149.5+2223, hereafter simply MACS~J1149,} at $z=0.54$ (Fig.~\ref{fig:stamp}). \citet{tripodi24} reported \id\ to have AGN characteristic features, including the line broadening in \hb, which supports the presence of a supermassive black hole of $10^8\,M_\odot$, as well as high-ionization UV lines. \id\ was reported to have blue UV and red optical continua, a characteristic spectral feature seen in the little red dot (LRD) population {\citep{matthee23,furtak23,labbe24c,barro24,greene24}}.
% {The broad component detected in the \hb\ line supports of the presence of an AGN, with the black hole mass of $\sim10^8\,M_\odot$ \citep{tripodi24}. 
% While line broadening could also be of non-AGN origin, such as outflow \citep[e.g.,][]{marques-chaves24}, the absence of a broad component in other lines (e.g., \oiiir\ of higher SNR) disfavors the outflow scenario. 
% In addition, the absence of high ionization lines ($>54$\,eV) in \id\ makes its nature as AGN even more intriguing. All the observed emission lines other than broad \hb\ could essentially be created by low-metallicity, massive stars too, whereas typical AGN are often associated with higher ionization lines \citep{berg21}. The same situation applies to some other LRDs reported in the literature \citep[e.g.,][]{taylor25}, whereas \citet{tang25} reported the detection of \nv\ in a LRD at $z=6.98$. 
% Without detection of the broad \hb\ line, \id\ could have been classified as a star-forming galaxy of high ionization.
% (``very high-ionization"). 
% The observed emission line property of \id\ is consistent with the ``high-ionization" class (35--55\,eV) of \citet{berg21}. 
% }

As we see below, a newly taken NIRSpec G140H/F070LP spectrum reveals broad \ly\ emission in \id. 
{The detection of \ly\ is considered rare at $z>7$ i.e. the epoch well before reionization completes, due to the high IGM neutral fraction \citep{Mason2018}. Despite, several studies reported \ly\ detection \citep[e.g.,][]{zitrin15,matthee18,larson22,roberts-borsani23}. In many cases \ly\ detection is found in relatively UV-bright galaxies $M_{\rm UV}<-20$\,mag, where the presence of a large ionized bubble is considered to assist \ly\ photons to escape \citep{cen00,dijkstra14}. Recent studies with sensitive JWST spectroscopy advanced our understanding of \ly\ emitters with a statistical sample of galaxies \citep[e.g.,][]{nakane24,witten24} and detailed line-profile analysis \citep[e.g.,][]{witstok25}.}
Our G140H spectrum of \id\ reveals the \ly\ emission near the systemic velocity, which, at that redshift, indicates the presence of a large ionizing bubble. With the sensitivity offered by the observatory, further assisted by lens magnification, in this study our aim to analyze the \ly\ line profile and characterize the ionizing structure around this early AGN.

% Our reanalysis of the existing PRISM spectrum reveals the potential enrichment of nitrogen, as represented by its enormously bright \niv\ line. 
% In addition to the great sensitivity offered by the observatory, the lens magnification by the foreground cluster allows us to study the properties \id\ with greater sensitivity. 

The paper is outlined as follows: In Sec.~\ref{sec:data}, we present our data reduction and analyses. In Sec.~\ref{sec:ana}, we analyze emission lines of \id. In Sec.~\ref{sec:disc}, we discuss the nature of \id\ in the context of recent discoveries of luminous, nitrogen-rich galaxies and AGN at $z>6$. Where relevant, we adopt the AB magnitude system \citep{oke83,fukugita96}, cosmological parameters of $\Omega_{\rm m}=0.3$, $\Omega_\Lambda=0.7$, $H_0=70\,\kms\, {\rm Mpc}^{-1}$, and the \citet{chabrier03} initial mass function (IMF). 

%%%%%%%%%%%%%%%%%%%%%
\section{Data}\label{sec:data}

%%%%%%%%%%%%%%%%%%%%%
%%%%%%%%%%%%%%%%%%%%

%%%%%%%%%%%%%%%%%%%%%
\begin{figure*}
\centering
	\includegraphics[width=0.9\textwidth]{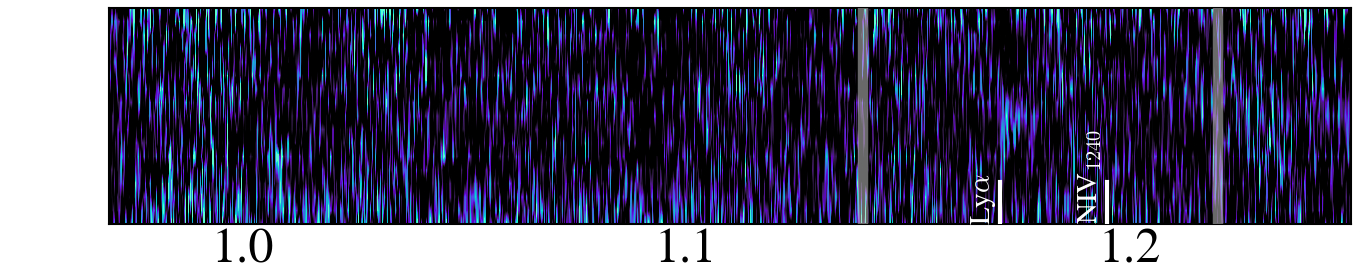}
	\includegraphics[width=0.9\textwidth]{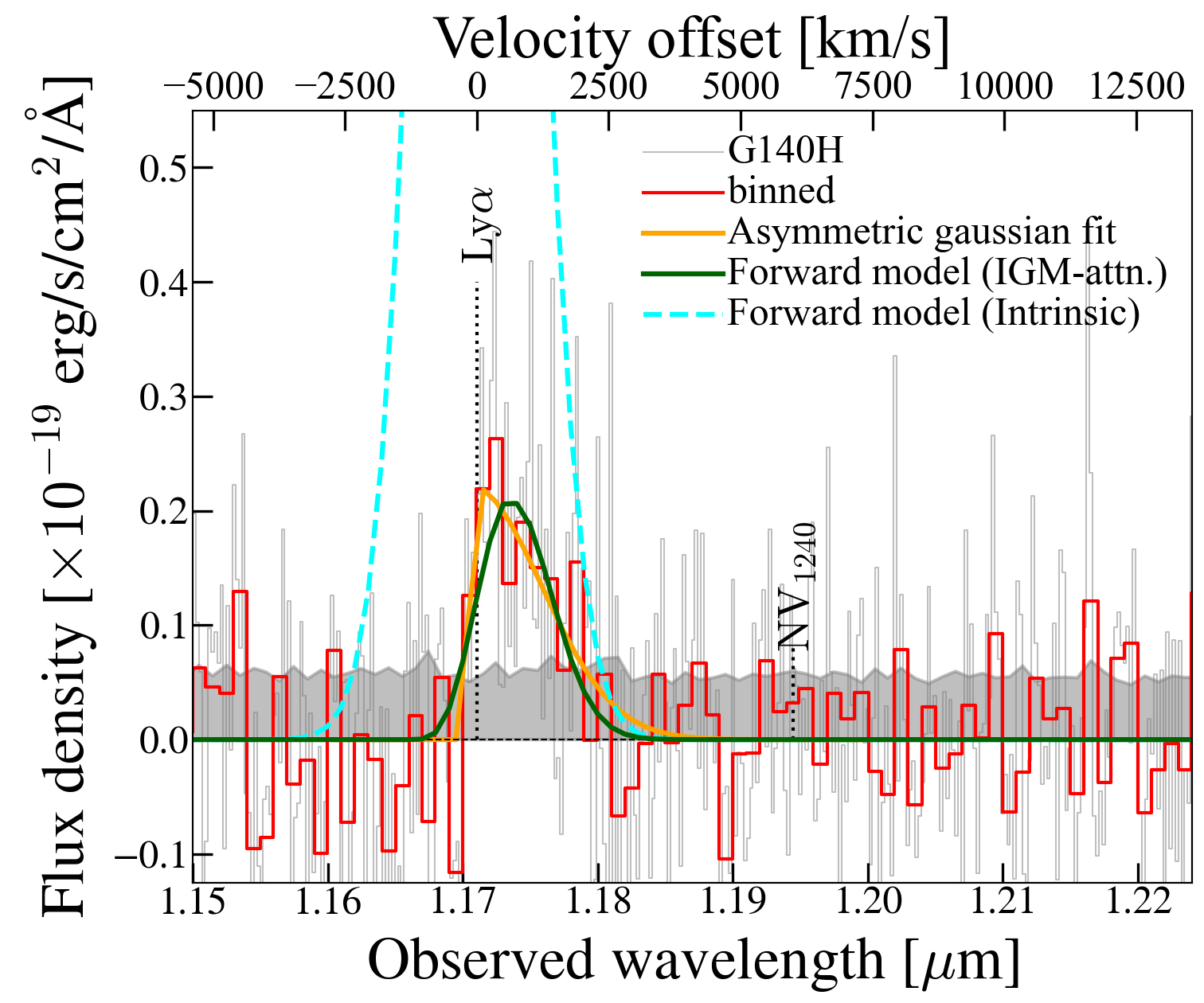}
	\caption{
    (Top): Two-dimensional NIRSpec G140H/F070LP spectrum of \id. The wavelength locations of \ly\ and \nv\ are indicated (red lines).
    (Bottom): One-dimensional spectrum, in the original (gray solid lines) and binned (red) pixel scales. {The forward-modeled \ly\ line profile (green solid line), and its intrinsic model (cyan dashed line), inferred from the IGM transmission modeling (Sec.~\ref{sec:emi_g140}), are shown. An asymmetric Gaussian model (Eq.~\ref{eq:1}) is also shown (orange solid line).}
    }
\label{fig:grating}
\end{figure*}
%%%%%%%%%%%%%%%%%%%%

%%%%%%%%%%%%%%%%%%%%%
\begin{figure}
\centering
	\includegraphics[width=0.48\textwidth]{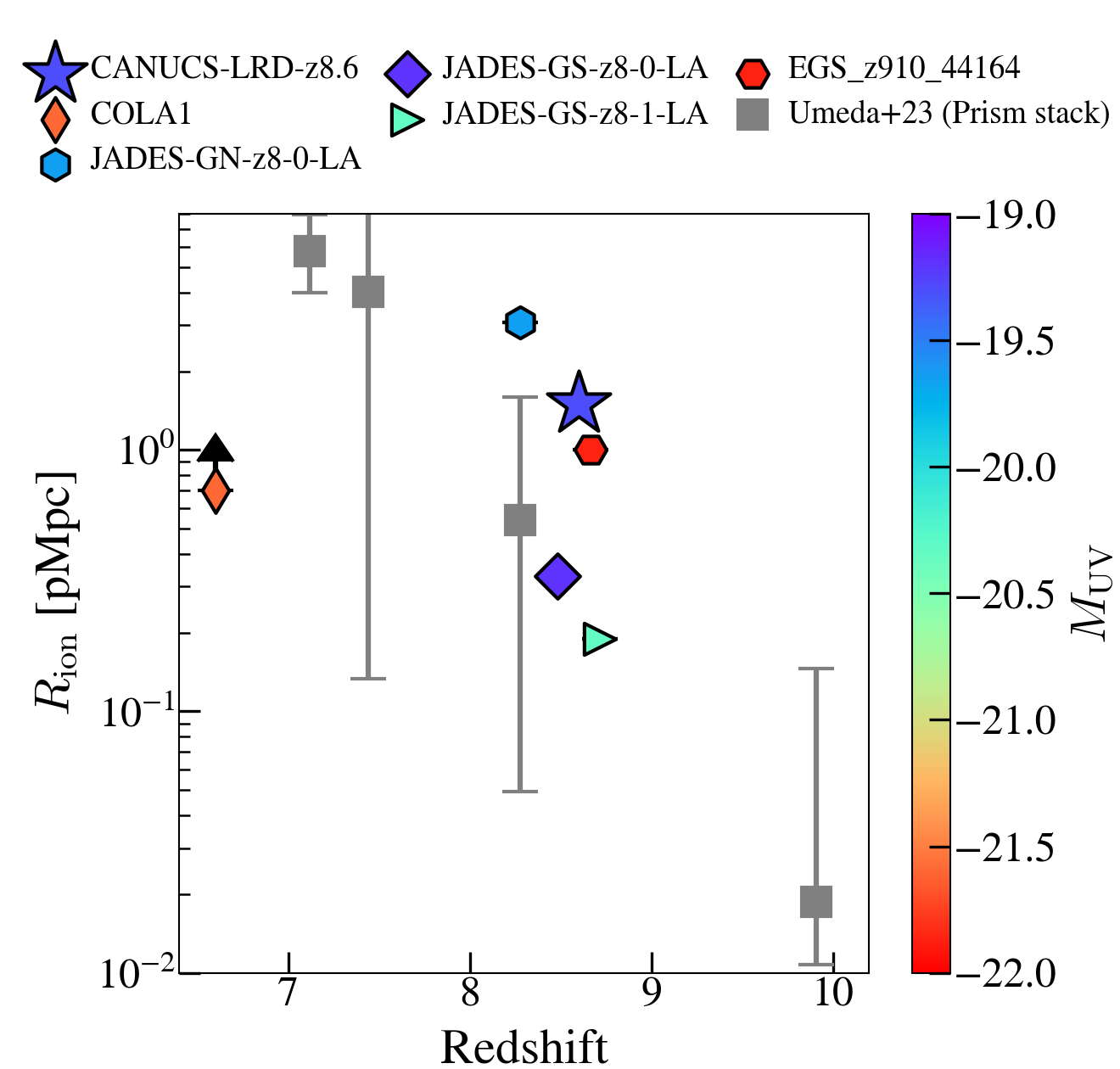}
	\caption{
    {
    Ionized bubble size distributions of \id\ and galaxies at $z>6$ in the literature. Sizes for COLA1 \citep{matthee18,mason20} and JADES-GN-z8-0-LA/JADES-GS-z8-0-LA/JADES-GS-z8-1-LA \citep{witstok25} were measured from analysis of the \ly\ line profile, similar to the one here. The size for EGS\_910\_44164 was inferred by the fact that it is located at $\approx1$\,Mpc away from another \ly\ emitter \citep{larson22}. \citet{umeda25} obtained the median bubble size of galaxies at each redshift through a \ly\ transmission and \ly\ damping wing absorption analysis performed on stacked NIRSpec/PRISM spectra.
    }
    }
\label{fig:zrb}
\end{figure}
%%%%%%%%%%%%%%%%%%%%

\subsection{NIRSpec/MSA G140H/F070LP observations}\label{sec:G140}
NIRSpec/MSA observations (GTO4552, PI Stiavelli) were configured with a high-resolution grating (G140H/F070LP), aiming to capture rest-frame UV emission lines of galaxies at $z\sim3$--9. The observations were executed on June 9 2025, pointed toward (R.A.,Decl.)=(11:49:38.16, +22:23:30.650) at position angle PA=258\,degree. The total science time, excluding overhead, is 14\,hrs, split into two MSA masks. A NIRCam parallel imaging (F140M/F430M, F070W/F360M) towards the 1199 parallel field \citep{morishita24,stiavelli25,zhang25} was attached to the primary MSA observations (Morishita, in prep.).

We designed two slit masks, following the same procedure as for the GTO1199 program \citep{stiavelli23,morishita24b}. The parent source catalog was constructed using existing HST imaging data, including CLASH, HFF, GLASS, and BUFFALO \citep{postman12,treu15,lotz17,kelly18,steinhardt20}, and existing JWST NIRCam imaging data \citep{stiavelli23,morishita24b}. Spectroscopic redshift measurements available in the literature were incorporated \citep[][]{grillo16,treu16,shipley18,schuldt24}. In our mask design, the highest priority was given to spectroscopically confirmed sources at $z>5$, including MACS1149-JD1 at $z=9.1$ \citep{hashimoto18,hoag19,stiavelli23,bradac24}, strong emitters \citep{morishita24b, stiavelli25}, and other newly identified sources from our preliminary analysis of the Cycle~1 PRISM data, including the main target of this study, \id. Because of spatial distributions of these high-priority sources, a large dither was implemented ($\sim0.\!'5$) between the two masks. For each source, the default 3 shutter slitlets were assigned, enabling exposure at three positions separated for $0.\!''53$ from each. We selected the default Entire Open Shutter Area for source allocations in APT, which secures placing the source center position within 35\,milliarcsec from the shutter boundary. Including fillers and open shutters in empty sky regions, our MSA masks consisted of in total 98 unique sources.

For the reduction of MSA data, we use {\tt msaexp}\footnote{https://github.com/gbrammer/msaexp} (ver0.9.5.dev8+ge2b237b), equipped with the official JWST pipeline (ver.1.18.0) and the pmap context 1364, following \citet{morishita24b}. The one-dimensional spectrum is extracted via optimal extraction. The source light profile along the cross-dispersion direction is measured directly using the 2-d spectrum. For sources with faint continuum or with any significant contamination (i.e. from a failed open shutter), we visually inspect the 2-d spectrum and manually define the extraction box along the trace where any emission lines are identified. 
% {The summary of the observed sources is presented in Appendix.} %%% Do this at the revision?

% The extracted 2-d and 1-d spectra of \id\ are shown in Fig.~\ref{fig:grating}.

%%%%%%%%%%%%%%%%%%%%%
\begin{figure*}
\centering
	\includegraphics[width=0.999\textwidth]{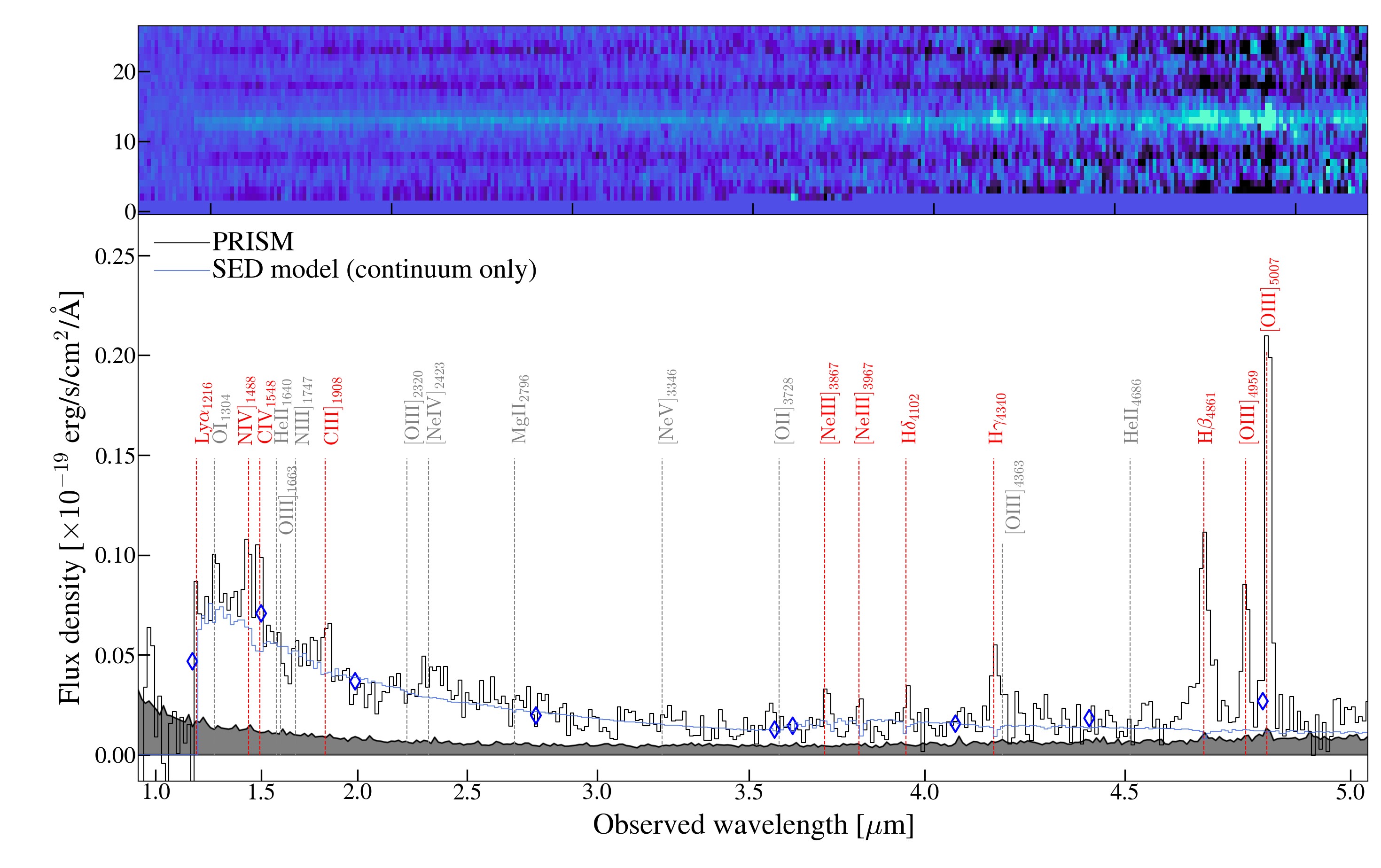}
	\caption{
    (Top): Two-dimensional PRISM spectrum of \id.
    (Bottom): One-dimensional PRISM spectrum of \id\ (black solid line) along with flux uncertainties (black shaded region). The best-fit SED model (stellar model only; blue line) used for continuum subtraction and NIRCam photometric data points (blue diamonds), to which the PRISM spectrum is normalized, are shown. Detected lines ($S/N>3$) are labeled in red, others in gray. \ly\ is also shown in Fig.~\ref{fig:grating}.
    }
\label{fig:prism}
\end{figure*}
%%%%%%%%%%%%%%%%%%%%

%%%%%%%%%%%%%%%%%
\subsection{NIRSpec/MSA Prism observations}\label{sec:prism}
We utilized the prism observations performed as part of GTO1208 (CANUCS; \citealt{willott22,sarrouh25}). The exposure time on \id\ (ID~5112687 in their MSA catalog) is 2845\,sec. For more details of the observations, interested readers are referred to their survey paper. The spectrum for \id\ is reduced in the same manner as for G140H spectra. Because the continuum flux is significantly detected, the extracted spectrum is then corrected for aperture loss by matching the pseudo-F150W flux to the corresponding NIRCam photometric flux. 
% The extracted spectrum is shown in Fig.~\ref{fig:prism}.

%%%%%%%%%%%%%%%%%%%
\subsection{Imaging Data Reduction and Photometry}\label{sec:imaging}
Imaging data are retrieved from the MAST archive, including filters F090W, F115W, F150W, F200W, F277W, F356W, F410M, F444W, and F480M, originally taken in the program 1208 \citep{willott22}, 1199 \citep{stiavelli23}, and 2883 \citep{fu25}. Those data were consistently reduced. Readers are referred to Morishita in prep. for details of the reduction procedure. All images were aligned and resampled in the pixel size of $0.\!''02$. 
{In addition, to increase the effective coverage for the search of photometric sources (Sec.~\ref{sec:ion}), we include HST/ACS images (F435W, F606W, F814W) available in the field from the BUFFALO project \citep{steinhardt20}.}

Source fluxes are measured on the PSF-matched (to the F444W point-spread function, PSF) images, with a fixed aperture of radius $r=0.\!''16$\,. The psf-matched fluxes are then scaled to the total flux by adopting a single scaling factor defined for each source by ${\rm flux_{total, F444W}/flux_{aper, F444W}}$. In the following analysis, we adopt the latest magnification model by \citet{schuldt24}{, one of the latest models publicly available. The model utilized 162 spectroscopically confirmed sources.} At the position of \id, we obtain the magnification factor $\mu=1.85_{-0.02}^{+0.02}${, which is consistent with other lens models from the Hubble Frontier Fields project ($\mu=1.82_{-0.10}^{+0.11}$). We note that \citet{tripodi24} reported a smaller magnification factor, $\mu=1.06$--1.15 using {\tt lenstool}. The cause of this discrepancy is unknown. However, the choice of the magnification does not change our main conclusions.}

%%%%%%%%%%%%%%%%%%%%%
\section{Analysis and Results}\label{sec:ana}

%%%%%%%%%%%%%%%%%%%%%
\subsection{Lyman-$\alpha$ line modeling in G140H/F070LP}\label{sec:emi_g140}
The G140H spectrum clearly detects the \ly\ line (Fig.~\ref{fig:grating}). The line has an extended faint envelope to the longer wavelength from the expected line position for the systemic redshift defined by $z_{\rm [OIII]5007}$.
% (see Sec.~\ref{sec:prism}).
This is naturally expected due to the resonant nature of \ly. The observed line flux peaks at $+3.6$\,\AA\ redward of the systemic \ly\ wavelength, or $\Delta v_{\rm red} \sim 880$\,km/s. 

{Since the observed line profile shows an asymmetric feature, we first fit the observed line with an asymmetric gaussian profile, by multiplying a skewness term, $S$, to gaussian $\mathcal{N}(\lambda_0, \sigma)$,
\begin{equation}\label{eq:1}
y = S(\lambda_0, \sigma, s) \mathcal{N}(\lambda_0, \sigma),
\end{equation}
where
\begin{equation}
S(\lambda_0, \sigma, s) = 1 + erf \left( s (\lambda - \lambda_0) / \sqrt{2}\sigma \right).
\end{equation}
From the fit, we find the line centroid $\Delta v_{\rm red} \sim128$\,km/s and FWHM $\sim1550$\,km/s (Fig.~\ref{fig:grating}). 
% The Gaussian fit to the observed line shows a larger offset, centered at $\Delta v_{\rm red} \sim 760$\,km/s. The observed line width is broad, FWHM\,$=1540\pm260$\,km/s. 
}

Remarkably, we observe \ly\ emission near the systemic line center, implying that the surroundings of \id\ is highly ionized --- the predicted transmission from the IGM damping wing at line center is $< 10\%$ if a source sits in an ionized region $< 0.5$\,pMpc, versus $>30\%$ if the ionized region is $>1$\,pMpc \citep{mason20}. In the observed spectrum, $\sim26\,\%$ of the \ly\ flux is found within $<250$\,km/s of the systemic velocity. To quantify the transmission structure, we follow the IGM prescription described in \citet{mason20}. We assume an intrinsic Gaussian \ly\ emission line, centered at the systemic velocity (see Sec.~\ref{sec:emi}), and include \ly\ optical depth $\tau_{\rm D}$ from a simple transmission model assuming a Str{\"o}mgren sphere of the size $R_b$ around the source i.e radius of the ionizing bubble.\footnote{\url{https://github.com/charlottenosam/LyaLineshapes}} 
We set the IGM neutral fraction $X_{\rm HI, IGM}=1$ outside the bubble and the residual neutral fraction inside the ionizing bubble $X_{\rm HI, ISM}=10^{-8}$, leaving $R_b$ as the only additional parameter. The IGM and ISM temperatures are set to $1\times10^{3}$\,K \citep[][]{hera23} and $1\times10^{4}$\,K, respectively, but changing these assumptions does not affect the final results. We note that the redshift is set as a free parameter within the $\pm1\,\sigma$ range of $z_{\rm [OIII]5007}$, to account for the relatively large uncertainty originating in the prism spectrum. Adopting different $X_{\rm HI, ISM}$ ($0$ and $10^{-4}$) does not change the main result.

The best-fit model is shown in Fig.~\ref{fig:grating}. We find the bubble size $R_b=1.5_{-0.2}^{+0.3}$\,pMpc. At $z\approx8.6$, the decline in \ly\ emission from most galaxies, and \ly\ damping wings, imply that the IGM is expected to be mostly neutral and with small ionized regions \citep[e.g.,][]{nakane24,tang2024,mason25}, making $>1$\,pMpc bubbles rare \citep{witstok25}, even in overdensities \citep{lu24a}. {The bubble size of \id\ is compared with other measurements of similar redshift galaxies in Fig.~\ref{fig:zrb}.} 
% The model peaks at $\Delta \lambda = +19.6$\,\AA\ off from the systemic redshift, corresponding to $\Delta v=+501$\,km/s. 
% We note that the observation of flux close to systemic velocity strongly favors a large ionized region regardless of the assumed line shape. 
If we instead assume the intrinsic line as a half-Gaussian, truncated at the systemic velocity, a much larger ionized bubble would be inferred, requiring almost 100\,\% transmission to match the observed line profile.
{We note that the goodness of fit, evaluated with the Bayesian Information Criterion (BIC), remains similar between the asymmetric Gaussian and IGM transmission models.}

The line width of the inferred intrinsic \ly\ line (cyan line in Fig.~\ref{fig:grating}) is found to be broad, with FWHM\,$\sim$\fwinterr\,km/s.
This is much broader than expected for typical star forming galaxies. In fact, as we see in the following (Sec.~\ref{sec:emi}), the broad \hb\ component has a similarly large velocity width, $\sim3600\pm500$\,km/s. 
% While these do not match exactly, the smaller width measurement for \ly\ could be attributed to the complexity of its scattering nature and the configuration adopted in our line profile modeling. Regardless, t
The broad \ly\ line width thus suggests that the line originates in the AGN broad-line region (BLR). The detection of broad \ly\ close to the systemic velocity implies a lower covering fraction of dense gas (and dust) around the BLR, which highlights the unique nature of \id\ compared to LRDs in the literature \citep[e.g.,][]{torralba25}. We note that our prism spectrum analysis shows no evidence of outflow. 

By comparing the observed and intrinsic line fluxes, we find the \ly\ escape fraction of \escerr\,\%, which is considerably high for the redshift \citep[cf.][]{morishita23b,nakane24}, and is rather close to those at lower redshift $z\sim5$--6 \citep[e.g.,][]{chen24,yue25}, implying low IGM opacity. Similarly, we can compare the intrinsic \ly\ flux to \hb. For the electron temperature of $3\times10^4$\,K and metallicity of $0.01\,Z_\odot$ from \citet{tripodi24}, and the electron density of $n_e=10^3\,{\rm cm^{-3}}$, the line ratio is expected to be $I_{1216}/I_{4861}\sim25$, calculated with {\tt Cloudy} photoionization code. Using the total \hb\ line flux, the observed line ratio is $I_{1216}/I_{4861}=3.0\pm0.8$, which translates to the escape fraction of $\sim12$\%, consistent with the one found above. The similarity found in the two \ly\ escape fractions attributes the attenuation of \ly\ to the IGM, suggesting that the ISM/CGM is highly ionized.

The G140H spectrum has wavelength coverage up to $\sim1.29\,\mu$m. Although \nv\ is within the coverage, the line was not detected.

%%%%%%%%%%%%%%%%%%%%%
\begin{figure}
\centering
	\includegraphics[width=0.48\textwidth]{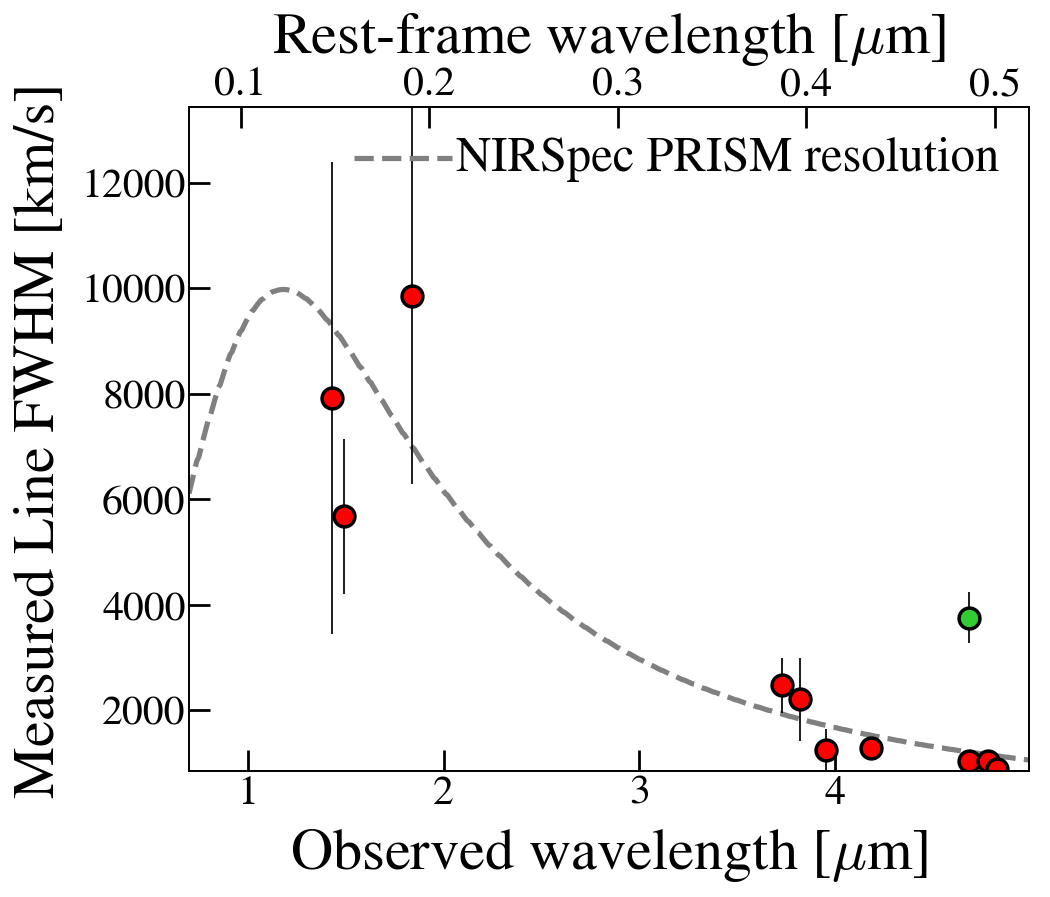}
	\caption{
    Width measurements of the lines detected in the prism spectrum ($S/N>3$; circles with error bars). The dashed curve compared is {a nominal resolution limit} for NIRSpec PRISM. All lines, except for the broad \hb\ component (green), are unresolved.
    }
\label{fig:linewidth}
\end{figure}
%%%%%%%%%%%%%%%%%%%%

% %%%%%%%%%%%%%%%%%%%%%
% \begin{figure*}
% \centering
%     \includegraphics[width=0.999\textwidth]{1208_5112687_yst5112687_fit_chakraborty.png}
% 	\caption{
%     Inference of the oxygen abundance, 12+\logoh, using strong line calibrators from \citet[][black lines]{chakraborty24}. The measurements are shown with $1\,\sigma$ uncertainties (red circles). In the bottom-right panel, \id\ is compared with $z>3$ galaxies in the mass-metallicity plane. The best-fit relation at $z\sim8$ is shown (dotted line; \citealt{morishita24b}).
%     }
% \label{fig:logoh}
% \end{figure*}
% %%%%%%%%%%%%%%%%%%%%

% %%%%%%%%%%%%%%%%%%%%%
\subsection{Emission Line Measurement in Prism}\label{sec:emi}
We model the line profile of each emission line of interest in the extracted 1d PRISM spectrum (Fig.~\ref{fig:prism}) with a Gaussian function. Our basic strategy is to define a wavelength window for each line, subtract the underlying continuum spectrum, and perform a Gaussian fitting. 
% For continuum subtraction, we utilize the best-fit stellar-only (excluding nebular component) SED template derived by {\gsf} (Fig.~\ref{fig:prism}; also see Sec.~\ref{sec:sed}). 
The line model includes the amplitude, line width, and redshift parameters. While most lines are unresolved in the Prism spectrum, we find significant line broadening around the \hb\ line, at the observed wavelength $4.68\,\mu$m, as also found by \citet{tripodi24}. We therefore add a broad line component at the same wavelength.

The flux of each line is estimated by integrating the corresponding Gaussian component, and the flux error is estimated by summing the error weighted by the amplitude of the Gaussian model in quadrature. The uncertainty associated with the modeled continuum is negligible but also added. In the following analysis, we adopt flux measurements when the line is detected (signal-to-noise ratio $S/N\geq3$); for those not detected, we adopt the 3-$\sigma$ upper limit. 

The measured line fluxes are reported in Table~\ref{tab:lines}. Lines detected above the signal-to-noise ratio $S/N$ greater than 3 are: \ly, \niv, \civ, \ciii, \neiii$_{\rm \lambda 3867,3967}$, \hd, \hg, \hb\ (narrow and broad), and \oiii$_{\rm \lambda 4959,5007}$. Other lines, such as \nv\, \oi, \heii$_{\lambda 1640}$, \oiiiuv, \nev, \oii$_{\lambda 3727}$, \oiii$_{\lambda 4363}$ (blended with \hg) are not detected. The Mg~\textsc{ii}$_{\lambda\lambda2797,2803}$ lines, which are characteristic to typical quasars, are not detected in the prism spectrum. The absence may be attributed to e.g., absorption by weakly ionized interstellar medium \citep[e.g.,][]{guseva13} but further details should be addressed with a higher resolution spectrum. 

{Overall, our line flux measurements are consistent with those reported in \citet{tripodi24}. \citet{tripodi24} reported flux measurements from three approaches. They detected \oiiif\ line at $S/N=2.7$--4.3, in all cases with the line width fixed to that of \oiiir, whereas our method leaves the width as a free parameter resulting in non-detection. Also, our \oiiir\ flux measurement is $\sim12$--$17\,\%$ larger despite the line being significantly detected. This is likely due to the difference in calibration files used in the two studies. In particular, the pathloss correction can have a significant effect near the wavelength edge of the NIRspec/PRISM sensitivity.}

% In addition, the prism spectrum firmly detects the \oi\ line (Fig.~\ref{fig:oi}), which was not reported in \citet{tripodi24}. The line complex (triplet of 1302\,\AA, 1304\,\AA, and 1306\,\AA) is often seen in AGN in emission but not in normal star-forming galaxies. The \oi\ line is believed to originate in the same region of broad line region clouds as [Fe~\textsc{ii}] \citep{rodriguez-ardila02}, which is the strongest coolant there. The physical origin is attributed mainly to the Bowen resonance fluorescence mechanism, with some additional contribution from collisional excitation \citep[e.g.,][]{matsuoka05}. 
% H~II Ly$\beta$ photons emitted by the ionized gas at rest wavelength 1025.72\,\AA\ can be absorbed by O~I at a nearly coincident wavelength (1025.76\,\AA), enhancing its $2p^3d^{3\ 3}D_3^{O}$ level. The level then quickly decays by a cascade of three fluorescent transitions in the NIR, optical, and UV, corresponding to the 11287\,\AA, 8446\,\AA, and 1305\,\AA\ lines. These lines are predicted to have equal intensities in photon units, giving the flux ratios of 3.4:2.0:0.7 --- while being brighter, the two red O~I lines are beyond the wavelength coverage by NIRSpec.

In our analysis in Sec.~\ref{sec:G140} and in the following sections, we rely on the redshift derived with the \hb+\oiii-doublet lines, $z=8.6329\pm0.0005$. This is slightly higher than that in \citet{tripodi24}, $z=8.6319\pm0.0005$, though consistent within the 1-$\sigma$ uncertainty. 
% The source of the discrepancy could be the different reduction algorithm and/or the pmap context used. 
Regardless, in this study we adopt the redshift measured using our reduced PRISM spectrum, to ensure the same reduction setup and line measuring as for the G140H spectrum.

% to discuss the relative velocity offset of \ly\ from the systemic redshift (Sec~\ref{sec:G140}).

In Fig.~\ref{fig:linewidth}, we compare the line width of the detected lines (FWHM, in km/s) with the theoretical limit for PRISM. Most lines detected in the prism spectrum here are consistent with the theoretical value within $1\,\sigma$ uncertainties, and thus we consider them unresolved. The only exception is the broad \hb\ component with FWHM\,$=3600\pm500$\,km/s, which supports the presence of AGN in \id. \cite{tripodi24} estimated its black hole mass to be $M_{\rm BH} = 1.0_{-0.4}^{+0.6} \times 10^8 M_\odot$ using a local empirical relation \citep{greene05}.
\section{Discussion}\label{sec:disc}

%%%%%%%%%%%%%%%%%%%%%
\begin{figure}
\centering
	\includegraphics[width=0.49\textwidth]{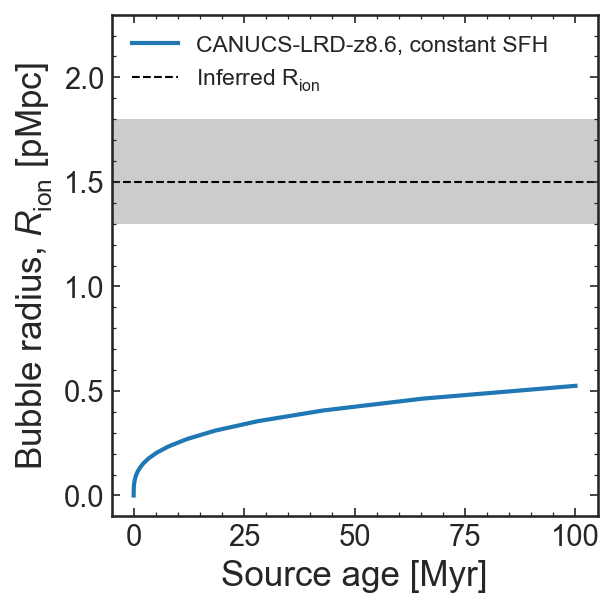}
	\caption{
    Ionizing bubble radius evolved as a function of time from a single ionizing source (blue line) compared to our inferred bubble size (gray shaded region). A constant ionizing source similar to \id\ is assumed. 
    }
\label{fig:bubble}
\end{figure}
%%%%%%%%%%%%%%%%%%%%

%%%%%%%%%%%%%%%%%%%%%
\begin{figure*}
\centering
	\includegraphics[width=0.5\textwidth]{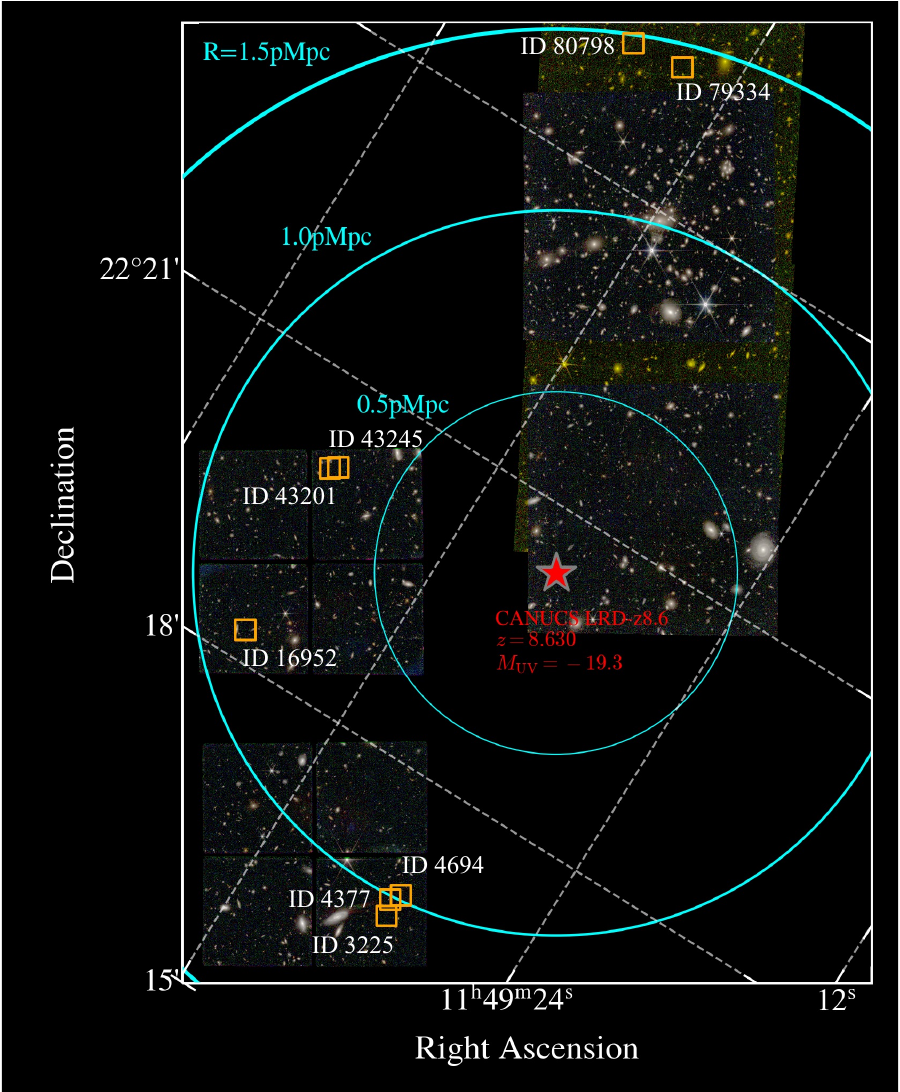}
	\includegraphics[width=0.49\textwidth]{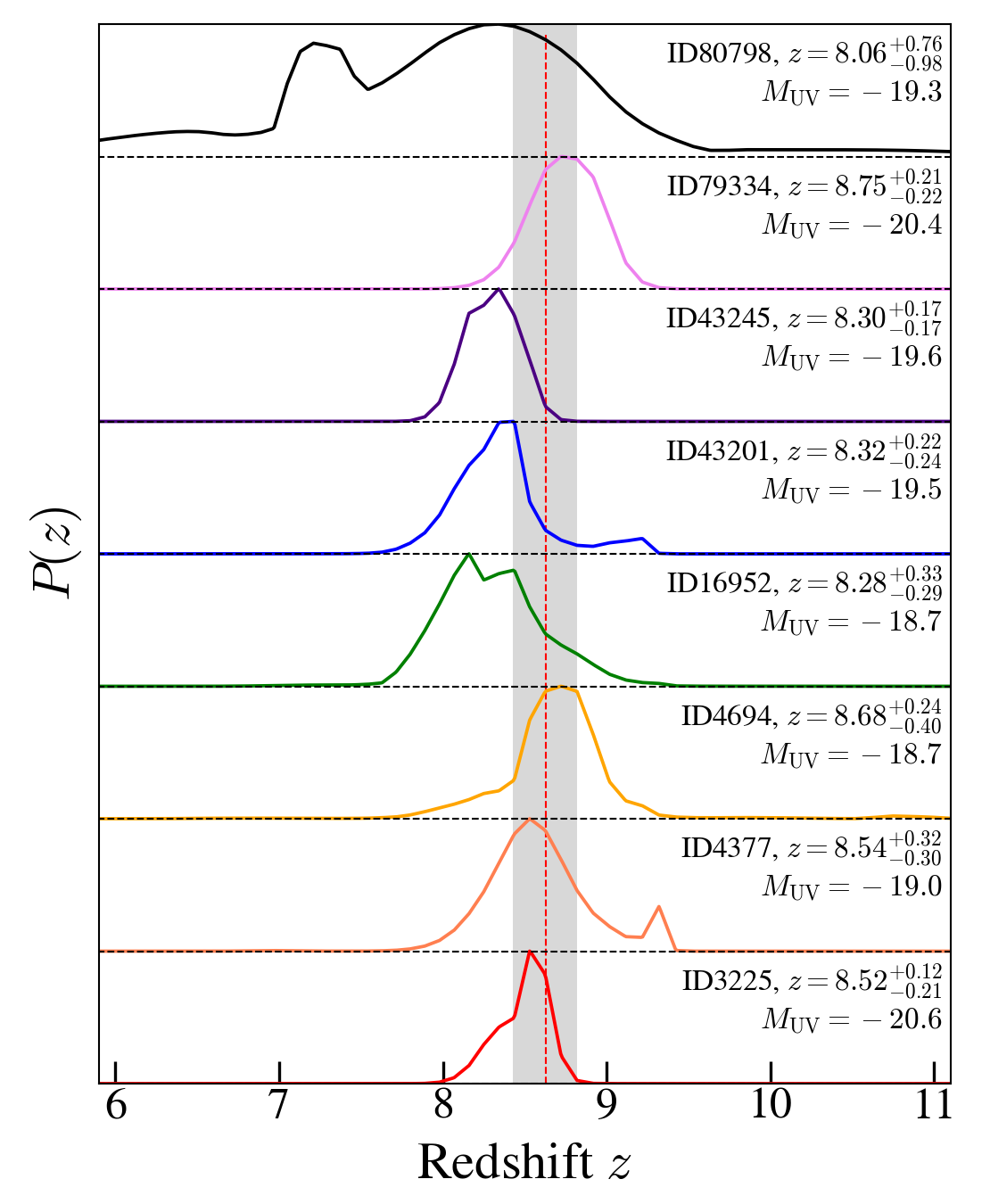}
	\caption{
    {
    ($Left$): Spatial distribution of potential member galaxies (orange squares) within an overdensity around \id\ (red star), shown on a pseudo RGB color image (NIRCam F090W/F115W/F150W for blue/green/red). \id\ is likely within a mild overdensity, with the overdensity factor of $\delta = $\,\dod\ (Sec.~\ref{sec:ion}). We note that the available NIRCam imaging coverage is incomplete within the projected bubble size of \id\ ($R_b\sim1.5$\,pMpc; Sec.~\ref{sec:emi_g140}). The image coverage shown represents the effective area for the photometric member candidate search.
    ($Right$): Photometric redshift probability distribution of individual candidate member galaxies.
    }
    }
\label{fig:radec}
\end{figure*}
%%%%%%%%%%%%%%%%%%%%

%%%%%%%%%%%%%%%%%%%%%%%%%%%%%%%%%%%%%%%%%%%%
\subsection{A Little Red Dot Residing in a Large Ionizing Bubble --- What Created It?}\label{sec:ion}
Lastly, the broad \ly\ emission observed in the G140H/F070LP spectrum supports the idea that a high-ionization environment enhances \ly\ photon escape. Our line-profile modeling supports the presence of a large ionized bubble, \rberr\,pMpc. We note that \id\ is not particularly UV bright ($M_{\rm UV}=-19.3$\,ABmag) cf. GN-z11 ($-21.5$\,ABmag) and GHZ2/GLASS-z12 ($-20.5$\,ABmag). It is not obvious whether such a large ionizing bubble was created by itself or whether there is a contribution from other nearby galaxies. 

Shown in Fig.~\ref{fig:bubble} is the time evolution of the ionizing bubble radius produced by $z=8.6$ assuming constant radiation from the source given its current UV luminosity, and accounting for recombination \citep[see e.g.][]{Shapiro87,mason20}. Even under this assumption, the maximum bubble radius that can be reached is $\sim0.5$\,pMpc at $100$\,Myr (i.e. the age of \id). This is much smaller than the one inferred from our analysis,
suggesting that a contribution from ionizing sources other than its own {\it star formation} is required. 
% We note assuming a constant ionizing output from \id\ is likely an upper limit, given the inferred SFH. 
A plausible additional ionizing source is photoionization by the central AGN. In this case, the ionizing photons might have been emitted at a specific cone angle that is aligned with the line of sight, creating a hole through which \ly\ photons may preferentially escape. 

In addition, we explored the spectroscopic data \citep{sarrouh25} and the photometric redshift catalog constructed here to assess if the region around \id\ is overdense, hosting other galaxies which could contribute to the formation of an ionized bubble. 
{To identify potential candidates, we first apply the Lyman-break selection method, following the same method described in \citet{morishita24beacon}. The selection requires non-detection at the wavelength shorter than rest-frame $\approx1216$\,\AA, where we require conservative SNR\,$<2$. For the redshift of interest, this corresponds to the NIRCam F090W and ACS F814W, F606W, and F435W filters. We further select robust candidates by applying photometric redshift cut. We exclude sources that do not satisfy both of the following ($i$) $8.43 < z < 8.83$ within the $1\,\sigma$ phot-z uncertainty, where the redshift range corresponds to $\sim\pm5$\,cMpc from \id, and ($ii$) $p(z>z_{\rm set})>0.8$ i.e. total redshift probability at $z>z_{\rm set}$ is greater than $80\,\%$, where we set $z_{\rm set}=5$. The combination of the non-detection requirement and the phot-$z$ selection provides us with robust sources whose phot-$z$ solution is consistent with the target redshift. Lastly, we visually inspect all selected sources, to exclude those with issues with flux estimates or any artifacts.
}
{From the photometric analysis, we find \nphot\ sources.} 
We find zero spectroscopic sources that are located within  $\Delta z < 0.05$ from \id. 

The spatial distribution of the photometric sources is shown in Fig.~\ref{fig:radec}. We note that the imaging coverage of the field occupies only a small fraction of the projected area of the inferred bubble radius, and thus a comprehensive investigation of its environment requires further data coverage. {While the HST imaging from the BUFFALO survey \citep{steinhardt20} offers additional coverage, the depth is much shallower for those areas than the NIRCam imaging used here. As such, additional analysis using only the HST data set would require a dedicated analysis, which is beyond our scope.}
Regardless, we can still estimate the overdensity factor, $\delta \equiv (n-\bar{n})/\bar{n}$, within the effective area covered by our NIRCam imaging. 

{To estimate $\bar{n}$, we run a completeness simulation as in \citet{morishita24beacon} and obtain effective volumes for our photometric selection. Briefly, we inject artificial sources of a fixed $M_{\rm UV}$ and $z$ in the NIRCam and ACS images. Each of those sources is assigned a single-slope SED randomly drawn from a set of various UV spectral slope, $\beta_{\rm UV}\in[{-}2.5{:}{-}1.5]$. We then detect these injected sources by following the same procedure of our photometry (Sec.~\ref{sec:imaging}) and run the Lyman-break selection as detailed above. We repeat this for the grids of $M_{UV}\in[{-}22{:}{-}16]$ and $z\in[7{:}10]$. The effective volume of each $M_{\rm UV}$ and $z$ is estimated by multiplying the recovery rate from our completeness simulation, $S(M_{\rm UV}, z)$. The volume is then corrected for magnification by dividing it by the magnification map. We note that the area where real sources are detected in the NIRCam imaging is also masked out in our completeness simulation and subtracted from the final effective volume estimate.}

{By using the volume estimated above and the luminosity function of \citet{mason22}, we expect $\bar{n}=$\,\nexp\ in average fields with our selection method, where the uncertainty is calculated by assuming the Poisson noise.} Using this number as a reference, we obtain {$\delta =$\,\dod}, suggesting a mild overdensity. Those photometric candidates, especially relatively bright ones at $M_{\rm UV}<-19$\,mag, are a potential contributor to the inferred ionizing bubble. {Notably, three candidates are comparably UV-bright as \id\ and two are $<-20$\,mag (right panel, Fig.~\ref{fig:radec}).} If these galaxies are indeed within the ionizing bubble of \id, non-zero \ly\ photon escape is expected in these galaxies too.

An interesting comparison can be made with A2744-ODz7p9, an overdensity of galaxies at $z=7.9$ that consists of 16 spec-$z$ confirmed members \citep{morishita23b,morishita25,hashimoto23,witten25}. Despite A2744-ODz7p9 being a lower-$z$, more significant overdensity ($\delta = 44_{-31}^{+89}$), only one of its members shows \ly\ emission \citep{chen24}. None of the member galaxies shows evidence of hosting an AGN.

%%%%%%%%%%%%%%%%%%%%%
\begin{figure*}
\centering
	\includegraphics[width=1\textwidth]{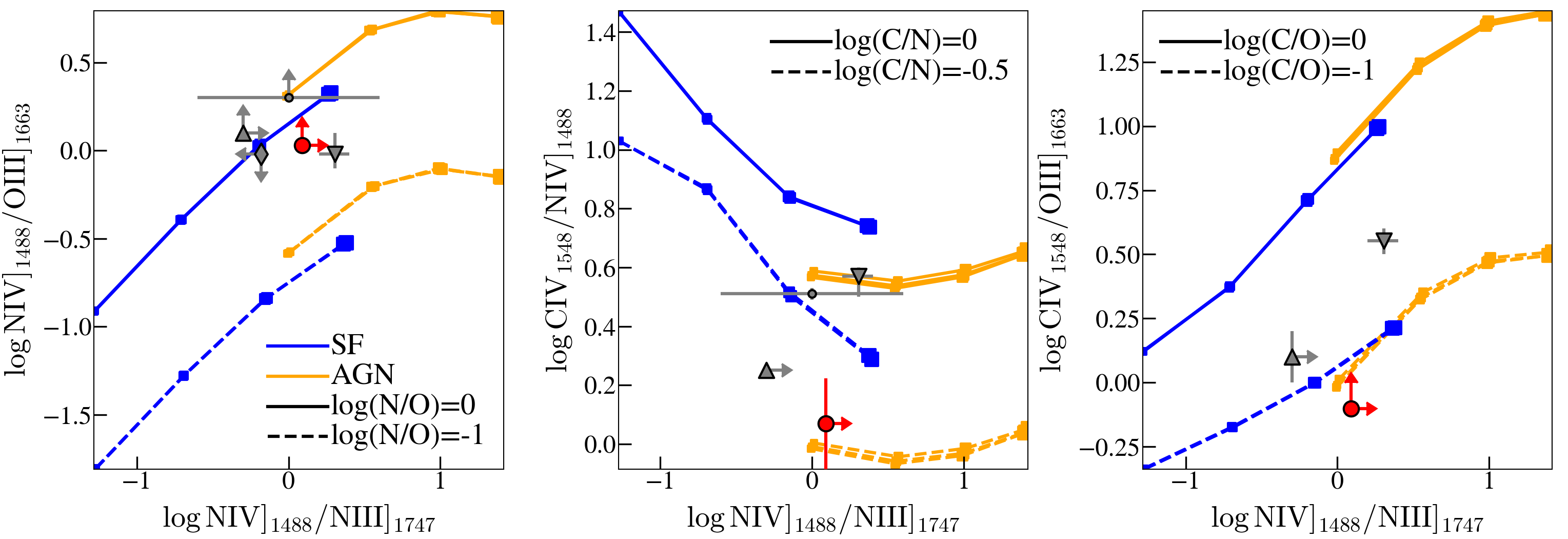}
	\caption{
    \id\ (red circles) is placed on rest-UV line diagrams. Four exotic objects in the literature are also shown for comparison: GHZ2 \citep[inverted triangle,][]{castellano24}, GN-z11 \citep[dot,][]{isobe23,maiolino24}, GLASS-150008 \citep[diamond,][]{isobe23}, CEERS-1019 \citep[triangle,][]{marques-chaves24}.
    (Left): Results from AGN (orange lines) and stellar (blue) photoionization models are shown. Models are calculated for different abundance ratios ($\log$\,(N/O)\,$=0$ and $-1$ with solid and dashed lines) and ionization ($\log U \in [-2.5:-1]$, scaled with the symbol size). Models are calculated with three densities ($n_{\rm H}=400,2000,20000$\,cm$^{-3}$; mostly overlapping each other).
    (Middle): Same as the left panel, but for \civ/\niv.
    (Right): Same as the left panel, but for \civ/\oiiiuv.
    }
\label{fig:diag}
\end{figure*}
%%%%%%%%%%%%%%%%%%%%

%%%%%%%%%%%%%%%%%%%%%
\subsection{\id\ in the context of Nitrogen-rich galaxies}\label{sec:n}
% \subsection{\id\ --- A Rare, Nitrogen-rich, Metal-Poor, Low-luminosity AGN}\label{sec:agn}
% In this section, we investigate the abundance of nitrogen over oxygen and carbon using line ratio diagnostics. 
% The black hole mass of \id\ (Sec.~\ref{sec:emi}) is $\sim1$\,dex overmassive compared to the local relation of \citet[][]{kormendy13}, but is within the range of those measured for LRD populations at $z>6$ in the literature \citep{harikane23c,kocevski23,akins25,maiolino25}. Strong X-ray emission is not detected in individual images of existing Chandra data \citep[e.g.,][]{ogrean16} at the source position. 

% The absence of high ionization lines ($>54$\,eV), on the other hand, makes the nature of \id\ as AGN intriguing. All the observed emission lines could essentially be created by low-metallicity, massive stars too, whereas typical AGN are often associated with higher ionization lines (``very high-ionization"). The observed emission line property of \id\ is consistent with the ``high-ionization" class (35--55\,eV) of \citet{berg21}. The same situation applies to some LRDs reported in the literature \citep[e.g.,][]{taylor25} (cf. \citealt{tang25} who reported the detection of \nv\ in a LRD at $z=6.98$). Without detection of the broad \hb\ line, \id\ could have been classified as a star-forming galaxy of high ionization.

% In addition, 
\id\ shows very strong \niv\ emission (Fig.~\ref{fig:prism}). {As detailed in \citet{tripodi24},} this line is very rare in low-$z$ {\it luminous} AGN ---  \citet{glikman07} found an increased fraction of \niv\ detection in their sample Type-I AGN at $z\sim4$, $\sim9\,\%$, from $\sim0\,\%$ in the local quasars. In fact, the UV \niv\ line has been reported in recent JWST studies of both broad-line \citep[e.g.,][]{ubler23,labbe24c,isobe25} and non-broad line sources \citep{maiolino24,topping24,isobe23b,marques-chaves24,castellano24}. The detection of strong \niv\ emission suggests an enhancement of nitrogen abundance in early AGN populations (see Sec.~\ref{sec:n}). 

%%%%%%%%%%%%%%%%%%%%%
% \id, with $M_{\rm UV}=-19.3$, is a low-luminosity AGN. In fact, the \niv\ line has been reported in recent JWST studies of both broad-line \citep[e.g.,][]{ubler23,labbe24c,isobe25} and non-broad line sources \citep{maiolino24,topping24,isobe23b,marques-chaves24,castellano24}. The detection of strong \niv\ emission suggests an enhancement of nitrogen abundance (see Sec.~\ref{sec:n}). 
% The detection of strong \niv\ emission suggests an enhancement of nitrogen abundance. 
The strong \niv\ emission hints at specific physical mechanisms in act, reflecting, e.g., anomaly seen during the CNO-cycles and strong wind from massive Wolf-Rayet stars \citep{kobayashi24,stiavelli25,zhang25}. In Fig.~\ref{fig:diag}, we show line flux measurements in three line diagnostics involving \niv, \nv, \oiiiuv, and \civ. For comparison, we calculated line ratios from photo-ionization models for AGN as follows: AGN models are calculated using {\tt Cloudy} photo-ionization code \citep{osterbrock89,chatzikos23}. We fix the gas-phase metallicity to $0.025\,Z_\odot$.
% one found in Sec.~\ref{sec:oh}. 
We calculate the model for different N/O and C/O abundance ratios ($\log$\,(N/O)\,$=-1$ and 0, $\log$\,(C/O)\,$=-0.5$ and 0) and ionization parameter ($\log U\in[-2.5:-1]$ in step of 0.5). For hydrogen density, we adopt $n_{\rm H}=400, 2000, 20000$\,cm$^{-3}$, to cover various densities reported for high-$z$ galaxies in the literature \citep[e.g.,][]{abdurrouf24}. For reference, we also calculate models for young star-forming galaxies. We use {\tt BPASS} \citep{eldridge17,stanway18} binary stellar radiation assuming an instantaneous star formation history with the stellar age of 10\,Myr, upper mass cut of 100\,$M_\odot$, and the Salpeter IMF.

From the comparison with the calculated model tracks, the observed \niv/\oiiiuv\ ratio of \id\ favors nitrogen enhancement, to $\log\,$(N/O)\,$\approx0$. Since \oiiiuv\ is not detected, its actual ratio can even be higher. The measured \civ/\niv\ suggests the enrichment of Nitrogen over carbon and/or under-abundant Carbon. The low \civ/\oiiiuv\ supports this too, but it remains unconstrained due to the non-detection of \oiiiuv. Also compared in Fig.~\ref{fig:diag} are exotic Nitrogen-rich objects reported in the literature ($\log$\,(N/O)\,$\simgt-0.5$), namely GHZ2 \citep[inverted triangle,][]{castellano24}, GN-z11 \citep[dot,][]{isobe23,maiolino24}, GLASS-150008 \citep[diamond,][]{isobe23}, CEERS-1019 \citep[triangle,][]{marques-chaves24}. \id\ is found in similar regimes occupied by those N-rich objects. A particular note is that the nature of those at $z>9.3$ remained largely undetermined, due to the lack of wavelength coverage for \hb\ \citep[see also][]{cleri25}. 
% The access to the \hb\ line for  hampered the solid confirmation of the presence of AGN in either case . 

% \citet{tang25} reported the detection of \nv\ in two objects, a galaxy at $z=8.7166$ and a LRD at $z=6.9827$, but \civ\ is not detected in either object, indicating nitrogen enhancement in a high-ionization region. \citet{topping24} found galaxy RXC~J2248-ID at $z=6.1$ shows strong \niv\ emission but not \nv. The authors found evidence of line broadening in \ha\ (FWHM\,$\sim600$\,km/s) {\it and} \oiii\ ($\sim1300$\,km/s). Also supported by the asymmetric line shape observed in \civ, the authors attributed the line broadening to an outflow origin. They concluded that the observed hard ionization is driven by massive star formation in very high density. 

One of the key common features of N-rich galaxies in the literature is high electron densities, ranging from $n_e\sim10^3$\,cm$^{-3}$ to $\simgt10^6$\,cm$^{-3}$. High electron densities are expected in both broad line regions and dense star-forming clouds. The electron density of \id\ could not be measured from the existing spectra, because neither of \niv- or \oii-doublets are resolved in the PRISM spectrum; \niv\ was beyond the wavelength coverage of G140H/F070LP. Alternatively, \id\ is characterized with a very compact morphology (Fig.~\ref{fig:stamp}). \citet{tripodi24} found it unresolved and placed an upper limit on its half-light radius, $<70$\,pc. 
% using {\tt galfit} \citep{peng02}, we find the effective radius $r_e=140\pm{10}$\,pc assuming the S\'ersic profile of $n=1$. 
This gives a star formation rate surface density $\Sigma_{\rm SFR, H\beta} > 3.5\times 10^2 M_\odot{\rm yr^{-1}\,kpc^{-2}}$ and a stellar mass surface density $\Sigma_* > 2.2\times 10^4\,M_\odot{\rm pc^{-2}}$ at the face value. These values are among the highest even in the literature of recent JWST findings \citep[e.g.,][]{morishita24}. \citet{schaerer24} indeed found that nitrogen emitters occupy the high $\Sigma_*$ and $\Sigma_{\rm SFR}$ regime, arguing that such peculiar conditions may enhance physical processes or ``exotic" sources of nucleosynthesis \citep[also][]{topping24,topping25}.

% An excellent reference sample at lower redshift is the well-studied Lynx arc at $z=3.36$ \citep{fosbury03,binette03}. \citet{fosbury03} found through photo-ionization modeling hot (80,000\,K) blackbody in a low metallicity ($0.038\,Z_\odot$) environment can produce the strong \niv, \oiiiuv, and \civ\ while \nv\ can be absent. A good candidate for the heating source is massive star formation, which is not unreasonable for \id\ given its redshift and observed oxygen abundance. The absence of \oiiiuv\ in \id\ may be explained by its lower oxygen abundance in ISM ($\sim0.025\,Z_\odot$) than the one adopted in the Lynx arc model.

% %=============================
\section{Summary}\label{sec:sum}
In this study, we presented a new JWST/NIRSpec observation of a previously known AGN, \id. We detected strong, broad \ly\ emission in the newly taken G140H spectrum. Considering the expected neutral fraction at the redshift ($X_{\rm HI}\sim1$), this suggests the presence of a large ionized bubble. Our line-profile fitting to the G140H spectrum indeed found $R_b=$\,\rberr\,pMpc using the prescription outlined in \citet{mason20}. The origin of this large ionizing bubble was discussed by investigating the photometric candidates in its surrounding. While a future followup is required, existing data indicate that \id\ is within a mild overdensity, {$\delta =$\,\dod}, suggesting that other galaxies surrounding might have contributed to the formation of the ionized bubble.

We also found exceptionally high \niv/\oiii\ ratio, revealed in our reanalysis of existing PRISM data. The detection of the very strong \niv\ line suggests an intriguing idea that \id\ may represent one of the evolutionary stages of Nitrogen-rich objects at $z>6$, recently discovered with JWST. Those N-rich objects are reported to also have high-ionization lines, but many of them are not classified as typical AGN either. In this regard, \id\ is a unique sample, as its nature is determined as AGN by the detection of broad \hb. A notable characteristic about \id\ is its high \niv/\niii\ ratio. One possible scenario is that, since \id\ hosts an AGN, it enhances the ionization of \niv, instead of \niii, in Nitrogen-enhanced ISM. The observed ratios of \niv, \civ, and \oiiiuv\ are found similar to those of nitrogen-enriched sources. Followup observations will determine its definitive abundances along with the electron density measurements. 
% The access to the \hb\ and \oiii\ lines at a decent $S/N$ level for \id, which are redshifted beyond the wavelength coverage by NIRSpec for objects at $z>9$, has thus given us a new insight into those peculiar N-rich objects.

%%%%%%%%%%%%%%%%%%%%%
% \begin{longrotatetable}
\begin{deluxetable*}{cccc}
\tablecolumns{4}
\tabletypesize{\footnotesize}
\tablewidth{-2pt}
\tablecaption{
Emission line measurements.
}
\tablehead{
\colhead{Line} & \colhead{Flux} & \colhead{FWHM} & \colhead{EW$_0$$^\dagger$}\\
\colhead{} & \colhead{$10^{-19}$\,erg/s/cm$^2$} & \colhead{km/s} & \colhead{$\mathrm{\AA}$}
}
\startdata
\cutinhead{G140H/F070LP}
${\rm Ly\alpha}_{1216}$ & $14.56 \pm 2.04$ & $1536.03 \pm 256.00$ & -- \\
${\rm Ly\alpha_{1216, intrinsic}}$ & $137.90 \pm 34.26$ & $2173.49 \pm 276.63$ & -- \\
${\rm NV}_{1240}$ & {$<5.4$}  & -- & -- \\
\cutinhead{PRISM/CLEAR}
${\rm OI}_{1304}$ & $<4.72$ & -- & --\\
${\rm NIV]}_{1488}$ & $16.66 \pm 5.00$ & $<7914.5$ & $19.1 \pm 7.6$\\
${\rm CIV}_{1548}$ & $19.57 \pm 3.70$ & $<5669.8$ & $27.0 \pm 5.6$\\
${\rm HeII}_{1640}$ & $<4.83$ & -- & --\\
${\rm OIII]}_{1663}$ & $<4.66$ & -- & --\\
${\rm NIII]}_{1747}$ & $<4.06$ & -- & --\\
${\rm CIII]}_{1908}$ & $18.08 \pm 4.24$ & $<9853.5$ & $36.8 \pm 15.8$\\
${\rm [OIII]}_{2320}$ & $<4.36$ & -- & --\\
${\rm [NeIV]}_{2423}$ & $<5.72$ & -- & --\\
${\rm [OII]}_{3727}$ & $<1.32$ & -- & --\\
${\rm [NeIII]}_{3869}$ & $8.98 \pm 1.12$ & $<2470.4$ & $55.6 \pm 11.2$\\
${\rm [NeIII]}_{3968}$ & $6.59 \pm 1.33$ & $<2200.9$ & $42.8 \pm 13.1$\\
${\rm H\gamma}_{4340}$ & $11.63 \pm 1.50$ & $<1274.4$ & $43.4 \pm 5.2$\\
${\rm [OIII]}_{4363}$ & $<1.59$ & -- & --\\
${\rm H\beta}_{\rm 4861, broad}$ & $26.45 \pm 3.86$ & $3554.9 \pm 480.4$ & $96.4 \pm 15.4$\\
${\rm H\beta}_{4861}$ & $20.04 \pm 2.27$ & $<1037.8$ & $78.5 \pm 11.7$\\
${\rm [OIII]}_{4959}$ & $19.19 \pm 2.18$ & $<1031.8$ & $67.9 \pm 6.2$\\
${\rm [OIII]}_{5007}$ & $57.79 \pm 2.77$ & $<874.1$ & $217.0 \pm 7.9$\\
\enddata
\tablecomments{
Fluxes are in units of $10^{-19}$\,erg/s/cm$^2$. Flux errors are $1\,\sigma$. $1\,\sigma$ upper limits are quoted for those undetected (S/N\,$<3$). 
$\dagger$: Rest-frame equivalent width.
}\label{tab:lines}
\end{deluxetable*}
% \end{longrotatetable}
% \input{table_phys.tex}

%=============================
\section*{Acknowledgements}
We thank the anonymous referee for a careful reading of the manuscript and for providing constructive comments.
The authors are grateful to Dan Coe, Martha Boyer, and Alison Vick for their support in planning the JWST program 4552. The authors are grateful to the CANUCS team for carefully planning and executing their observations. TM would like to thank Lee Armus for a useful discussion. Some/all of the data presented in this paper were obtained from the Mikulski Archive for Space Telescopes (MAST) at the Space Telescope Science Institute. The specific observations analyzed can be accessed via \dataset[10.17909/wct2-ga49]{https://doi.org/10.17909/wct2-ga49}. We acknowledge support for this work under NASA grant 80NSSC22K1294. TM received support from NASA through the STScI grants HST-GO-17231 and JWST-GO-3990. CAM acknowledges support by the European Union ERC grant RISES (101163035), Carlsberg Foundation (CF22-1322), and VILLUM FONDEN (37459). Views and opinions expressed are those of the author(s) only and do not necessarily reflect those of the European Union or the European Research Council. Neither the European Union nor the granting authority can be held responsible for them. SS has received funding from the European Union’s Horizon 2022 research and innovation programme under the Marie Skłodowska-Curie grant agreement No 101105167 - FASTIDIoUS.

{
{\it Software:} 
Astropy \citep{astropy13,astropy18,astropy22}, gsf \citep{morishita19},
numpy \citep{numpy}, python-fsps \citep{foreman14}, JWST pipeline \citep{jwst}.
}

%% For this sample we use BibTeX plus aasjournals.bst to generate the
%% the bibliography. The sample631.bib file was populated from ADS. To
%% get the citations to show in the compiled file do the following:
%%
%% pdflatex sample631.tex
%% bibtext sample631
%% pdflatex sample631.tex
%% pdflatex sample631.tex

\bibliography{output}{}
\bibliographystyle{aasjournal}

%% This command is needed to show the entire author+affiliation list when
%% the collaboration and author truncation commands are used.  It has to
%% go at the end of the manuscript.
%\allauthors

%% Include this line if you are using the \added, \replaced, \deleted
%% commands to see a summary list of all changes at the end of the article.
%\listofchanges

\end{document}